# GNC CHALLENGES AND OPPORTUNITIES OF CUBESAT SCIENCE MISSIONS DEPLOYED FROM THE LUNAR GATEWAY

Himangshu Kalita,[*] Miguel Donayre,[†] Victor Padilla,[‡] Anthony Riley,[§] Jesse Samitas,[**] Brandon Burnett,[††] Erik Asphaug,[‡‡] Mark Robinson,[§§] and Jekan Thangavelautham[***]

The Lunar Gateway is expected to be positioned on-orbit around the Moon or in a Halo orbit at the L2 Lagrange point. The proposed Lunar Gateway is a game-changer for enabling new, high-priority lunar science utilizing CubeSats and presents a refreshing new opportunity for utilization of these small spacecraft as explorers. In context, CubeSats are being stretched to their limits as interplanetary explorers. The main technological hurdles include high-bandwidth communications and reliable high delta-v propulsion. Advances in deep-space attitude determination and control has been made possible from the recent NASA JPL MarCO missions. Due to these limitations, CubeSats are primarily designed to be dropped-off from a larger mission. The limited mass and volume have required compromises of the onboard science instruments, longer wait times to send back science data to Earth, shorter mission durations or higher accepted risk. With the Lunar Gateway being planned to be closer to the Moon, it will provide significant savings for a propulsion system and provide a primary relay for communication apart from the DSN and enable tele-operated command/control. These three factors can simplify the mission enabling routine deployment of CubeSats into lunar orbit and enable surface missions. In this paper, we present preliminary designs of 2 CubeSat lunar landers that will explore the lunar pits, Mare Tranquilitatis and the remnant magnetic fields Reiner Gamma.

## INTRODUCTION

The proposed Lunar Gateway (Figure 1) will play an important role as a forward refueling base/pit stop on a journey to the Moon. Current plans are to position the base well outside the gravity wells of Earth and the Moon. A forward base such as the Lunar Gateway is needed to serve as a site for repairs/logistics hub and refueling depot in between any long journey to the Moon and beyond. This contrasts with Zubrin's Moon Direct approach[1]. While it is true the Apollo Program didn't require a hub or a forward base, it made the program all the more expensive and daring, that it couldn't be repeated for over 50+ years. A better solution is needed that provides a stepping stone to the Moon as part of a long-term, sustainable, two-prong strategy of science exploration and human colonization.

---

[*] PhD Candidate, Aerospace and Mechanical Engineering, Univ. of Arizona, 1130 N Mountain Ave., Tucson.
[†] Undergraduate Student, Aerospace and Mechanical Engineering, Univ. of Arizona, 1130 N Mountain Ave., Tucson.
[‡] Undergraduate Student, Aerospace and Mechanical Engineering, Univ. of Arizona, 1130 N Mountain Ave., Tucson.
[§] Undergraduate Student, Aerospace and Mechanical Engineering, Univ. of Arizona, 1130 N Mountain Ave., Tucson.
[**] Undergraduate Student, Aerospace and Mechanical Engineering, Univ. of Arizona, 1130 N Mountain Ave., Tucson.
[††] Undergraduate Student, Aerospace and Mechanical Engineering, Univ. of Arizona, 1130 N Mountain Ave., Tucson.
[‡‡] Professor, Lunar and Planetary Laboratory, Univ. of Arizona, 1629 E University Blvd, Tucson.
[§§] Professor, School of Earth and Space Exploration, Arizona State University, 781 E Terrace Rd., Tempe
[***] Assistant Professor, Aerospace and Mechanical Engineering, Univ. of Arizona, 1130 N Mountain Ave., Tucson.

Long term human colonization of the Moon requires further exploration of the lunar surface to better understand the origin, the formation and its composition in ever more detail. Significant insight has been obtained from the orbiting Lunar Reconnaissance Orbiter (LRO), however complementary surface missions are required to further our insight. Current plans for lunar science mission are all relatively expensive as they need to account for the cost of launch, upper-stage and the spacecraft. Use of the Lunar Gateway as a launch-pad for lunar flyby, orbit and surface mission can significantly cut down costs. The miniaturization of spacecraft electronics, sensors and actuators and recent demonstration of the MarCO Mars CubeSats[2] show that CubeSats despite being at their infancy as interplanetary explorers could be the ideal platform/vehicles for these flyby, orbit and surface missions to the Moon [3].

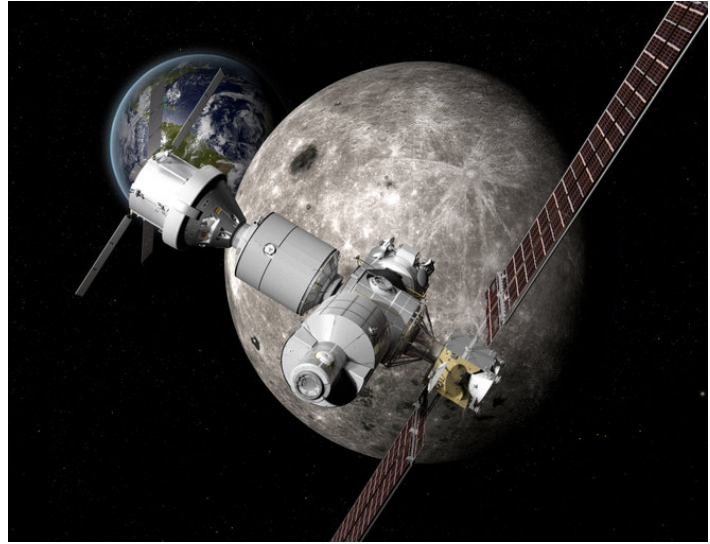

**Figure 1:** Current concepts of the Lunar Gateway (courtesy of NASA).

Compared to the SLS EM1 mission, the Lunar Gateway can play an important role not just a drop-off point but also as a permanent communication relay/hub to the CubeSats operating on the Moon. It may even be worthwhile to control and monitor the CubeSats from the Gateway rather Earth to ease communication congestion of the DSN. The relatively short distances could enable teleoperation of these CubeSats as opposed to fully autonomous operations. This could further simplify operation and minimize risks/uncertainties. The Lunar Gateway could also be used to insert an orbiting lunar relay and global positioning system to further facilitate exploration of both the far-side and near side of the Moon. Importantly these assets can increase localization accuracy to 10s of cm position accuracy and communication from with far-side assets. The relay and GPS system could be composed of CubeSats and such mission will further advance miniaturized ADCS systems for use in deep space.

Just as sending CubeSats to Low Earth Orbit (LEO) has become routine, it will be possible to do the same to the lunar vicinity. Sending these CubeSats to perform critical science exploration avoids the costs and risks of sending humans. The end to end mission costs could be reduced to $20 million and below for 54 kg, 27U CubeSats. Secondarily, the CubeSats being disposal can be used to further advance low-cost technology to navigate and precision land on the lunar surface, while advancing critical technologies like propulsion and communication which remain important technological hurdles[4,5,6]. The platform and the location maybe the ideal proving ground to test next-generation hybrid propulsion technologies that enables the CubeSats to perform sample-return from the lunar surface. Once lunar surface missions become routine, it may be possi-

ble to tackle Planetary Science Decadal questions including assembling telescopes and sensor networks on the far-side of the Moon[7].

In this paper, we present examples of lunar CubeSats that could be deployed from the Lunar gateway to perform exciting surface science. Preliminary design of two Lunar CubeSat lander missions called Arne II[8] and Cheesy Logic that would be deployed from the Lunar Gateway. One mission concept will be used to explore the lunar pits such as Mare Tranquilitatis and the second mission concept would be to deploy a lander to Reiner Gamma. Each lander is a 27U and has mass of 54 kg and each mission is expected to last no longer than 12 days. This avoids the thermal challenges of surviving the lunar night and simplifies the spacecraft design. Each lander will be self-propelled utilizing green-monoprop or solid-rockets with a maximum delta-v of 2.5 km/s. Arne II requires precision landing and will be required to land within 500 meters of Mare Tranquilitatis. In contrast, Cheesy Logic will need to land within the Renner Gamma region and have an accuracy 2 km. Through these mission, significant insight will be learned of lunar geology and geo-history. In addition, both missions will provide insight into developing a future human base.

In the following section, we will provide science motivations for the two mission concepts followed by related work, presentation of the spacecraft and concept of operations for each mission. This will be followed by detailed presentation of the mission trajectories and discussion, followed by conclusions and future work.

## BACKGROUND: SCIENCE MOTIVATIONS AND GNC CHALLENGES

### Remnant Magnetic Fields

The moon does not have magnetic dipoles like on Earth, however, it does have weak localized magnetic fields in Reiner Gamma, Mare Ingenii, and Mare Marginis. By studying these phenomena directly on the moon, we may be able to better understand if this is crustal or due to large impact events on the lunar surface that may be causing the magnetic fields[9,10,11].

Previous lunar missions have indicated that the swirl region has an elevated magnetic field in Reiner Gamma[9,10,11]. The readings were taken from a stable low orbit. It may be possible that an event hundreds of millions of years ago modified the magnetic properties of the surface materials, deflecting the solar wind and changing how the reflectance is modified by space weathering[12]. In Figure 2, a 3D simulation of the proton flux (a) and the magnetic flux (b) of Reiner Gamma. With the quantitative data that was simulated and with previous flights, TCL will be able to test one of several competing hypotheses trying to explain this science phenomena.

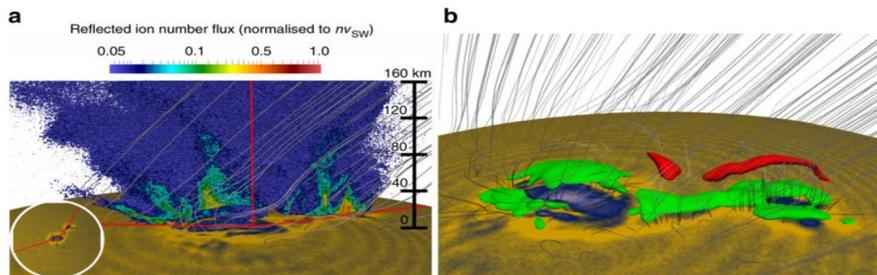

**Figure 2:** Magnetic Field Simulation Models of Reiner Gamma[12].

Landing on Reiner Gamma presents some important opportunities for visual navigation. The Reinner Gamma regions is visual distinct from the neighborhood. The suspected surface weathering has increased albedo of the lunar surface. Thus, simple visual navigation techniques could be used to identify, navigate and lock-on to the target.

**Lunar Pits**

Recently discovered lunar mare "pits" are key science and exploration targets. The first three pits were discovered within Selene observations and were proposed to represent collapses into extant lava tubes [13,14]. Subsequent Lunar Reconnaissance Orbiter Camera (LROC) images revealed 5 new mare pits and showed that the Mare Tranquilitatis Pit (MTP; 8.335°N, 33.222°E) opens into a sublunarean void at least 20 meters in extent (Figure 3) [14,15]. The pit diameters range from 86 to 100m with a maximum depth from shadow measures of ~107m. Several large, angular blocks are sparsely distributed across the floor, and likely represent detritus from the pit walls or collapsed roof materials.

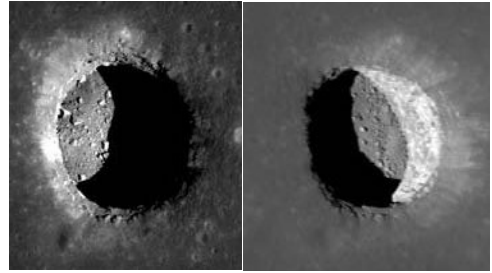

Current theories suggest there are vast networks of lava-tubes stretching 100s km. The suspected lunar lava tubes are expected to be much larger in size than those found on Earth, with a diameter of 80 to 100 m. Several of these lunar pits are found near the poles and conditions are sufficient to harbor water-ice. Entering and exploring these lunar pits are of high-priority. The lunar pits being sheltered from the surface could hold records of the lunar geo-history without being impacted by cosmic radiation, surface weathering and from micro-meteorites impacts. Furthermore, the temperature inside the pits are expected to be around -25 °C. These factors make the pits idea locations to locate a future human base or park valuable lunar assets.

**Figure 3:** Mare Tranquilitatis pit. (Left) near-nadir image and (Right) Oblique view (26° emission angle), a significant portion of the illuminated area is beneath overhanging mare.

Landing a CubeSat in the vicinity of a lunar pit presents some important GNC challenge. The main challenge includes precision landing near or inside the pit. This will require visual navigation techniques. One possibility includes looking for shadow lines inside the pit and lock on the target. Landing inside the pit presents some exciting outcomes but can serve to limit the length of the mission. Other options include landing nearby the pit and deploying smaller SphereX (Pit Bots) into the pit.

**RELATED WORK**

CubeSats and small satellites offer a new low-cost option to perform interplanetary exploration. First, we review mission concepts and missions that include an independent propulsion system or is deployed on an Earth escape trajectory. NASA JPL's INSPIRE is one such attempt that will result in a pair of CubeSats dropped off in an Earth escape trajectory[16]. INSPIRE is a pair of CubeSats that will fly past the moon to perform a technical demonstration. It includes a magnetometer, a deep space X-band communication system, computer and electronics. INSPIRE currently is in storage awaiting a launch opportunity. NASA JPL's MarCO CubeSats were intended to be communication relays for the Insight mission. The CubeSats were equipped with a DSN compatible x-band radio and tracking system, with a cold-gas propulsion system to perform minimal trajectory correction maneuvers. The mission was successful in acting as a communication relay and being a platform to test CubeSat components in deep space.

Another proposed interplanetary spacecraft concept is the Hummingbird jointly proposed by NASA Ames and Microcosm[17]. Hummingbird is a spacecraft architecture intended to tour asteroids. It includes slots to carry CubeSats that would be deployed upon rendezvous with a target of interest, in-addition, it includes a telescope to observe an asteroid target at a distance. Another interplanetary CubeSat is LunaH-Map, a 6U CubeSat selected for a NASA SLS EM1 mission[18].

LunaH-Map is a science focused mission using an experimental miniature Neutron Spectrometer to map speculated water ice deposits in the permanently shadowed craters of the Lunar South Pole. Lunar Ice Cube is a similar mission that will use an Infrared Spectrometer to look for water ice in the Lunar South Pole[19]. Lunar Flashlight is a third mission to explore the Lunar South Pole for ice deposit[20]. Another mission concept called Swirl will perform low-altitude orbits over Reiner Gamma to provide details maps of the remnant magnetic field[21]. It would use laser spectroscopy to identify presence of water ice. NEAScout is a proposed SLS EM1 mission intended to explore Near Earth Asteroids[22].

The most similar CubeSat mission to the ones presented is the OMOTENASHI CubeSat lander[23]. This CubeSat is a 6U, 14 kg lander that uses a hard landing system to land on the moon and will fly onboard SLS EM 1. OMOTENASHI has an airbag and crumple zone setup for the lander, whereas the proposed lander must use a soft-landing system because of the instruments on board. Another mission called LUMIO will travel to the L2 Lagrange point and monitor the moon for meteor impacts[24].

**SPACECRAFT DESIGN**

**Lunar Pit Exploration**

The proposed lander is a 27U (34 × 35 × 36 cm) CubeSat (Figure 4, 5). The lander will be deployed from the Gateway to the lunar surface. The lander is equipped with four 22N HPGP and four 100 mN HPGP thrusters. The High-Performance Green Propulsion (HPGP) provides higher specific impulse and higher propellant density, which results in increased performance compared to traditional propulsion. The propellant is based on AND (Ammonium DiNitramide) and is considered less toxic, non-carcinogenic and simpler to handle than hydrazine. The onboard Attitude Determination Control System (ADCS) consists of the Blue Canyon Technologies XACT-50. Reaction wheel desaturation will be performed using the thrusters.

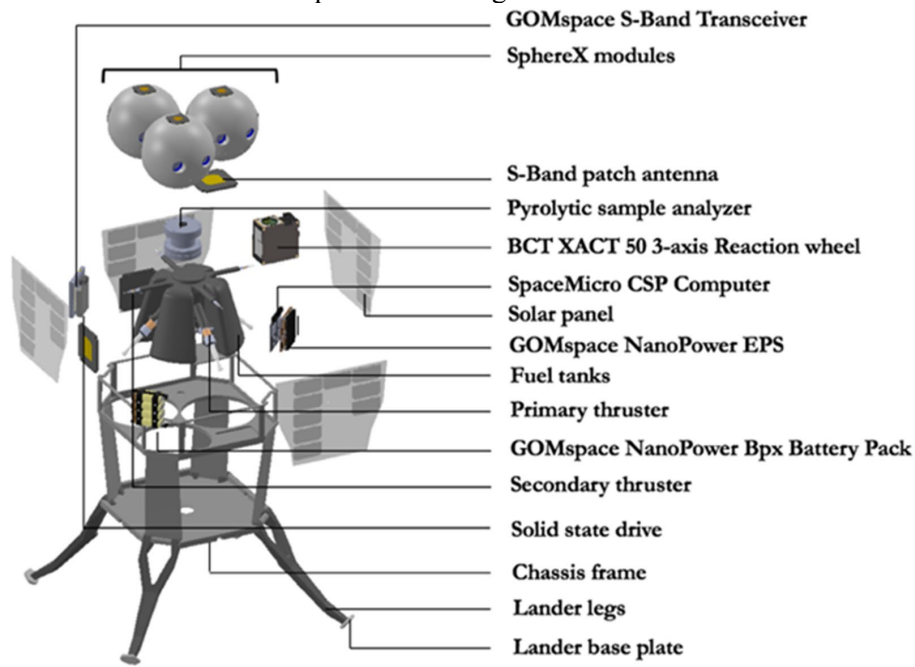

**Figure 4:** Lander for Lunar Pit Exploration

The power system consists of GOMspace NanoPower BPX rechargeable lithium ion batteries, MMA eHaWK solar PV and the power electronics from GOMspace NanoPower p60 system. The lander will also use 3 S-band antennas and a S-band transceiver. The C&DH will be performed by Space Micro CSP (CubeSat Space Processor) which is a compact single board computer designed around Xilinx Zync-7020. The chassis and the landing legs will be custom built. The instruments in the lander includes a pyrolytic analyzer (pyrolysis oven coupled to a mass spectrometer) for vacuum pyrolysis of regolith to identify the presence of water, and a camera and a lens system integrated with Xilinx Vertex-5QV FPGA for navigation during landing phase. The lander will also carry three spherical hopping robots (SphereX[25,26]) which will be deployed near the entrance of a lava tube. The SphereX robots will hop inside the lava tube for mapping, detecting the presence of water and collecting samples. The robots will perform autonomous navigation inside the lava-tube[30].

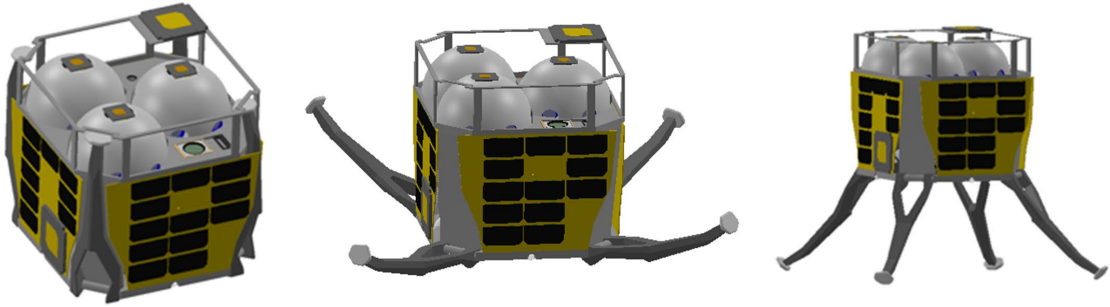

**Figure 5:** Landing legs deployment sequence.

**SphereX Hopping Robots**

SphereX[25,26] is a spherical hopping robot of mass 1.5 kg and a diameter of 180 mm as shown in Figure 6. The lander will carry three SphereX robots and deploy them near the entrance of a pit (Mare Tranquilitatis). The SphereX robots hops inside the pit for exploration. The top half of the SphereX contains lithium ion batteries and EPS board for power and EnduroSat S-band patch antenna for communication. The middle section consists of the C&DH board, 3-axis reaction wheel from Aerospace Corporation for attitude control, a pair of Bluefox 3 FPGA cameras for imaging and navigation and a 3D LiDAR scanner for mapping, navigation and localization[27]. The bottom half consists of a 5 N thruster along and a propellant tank.

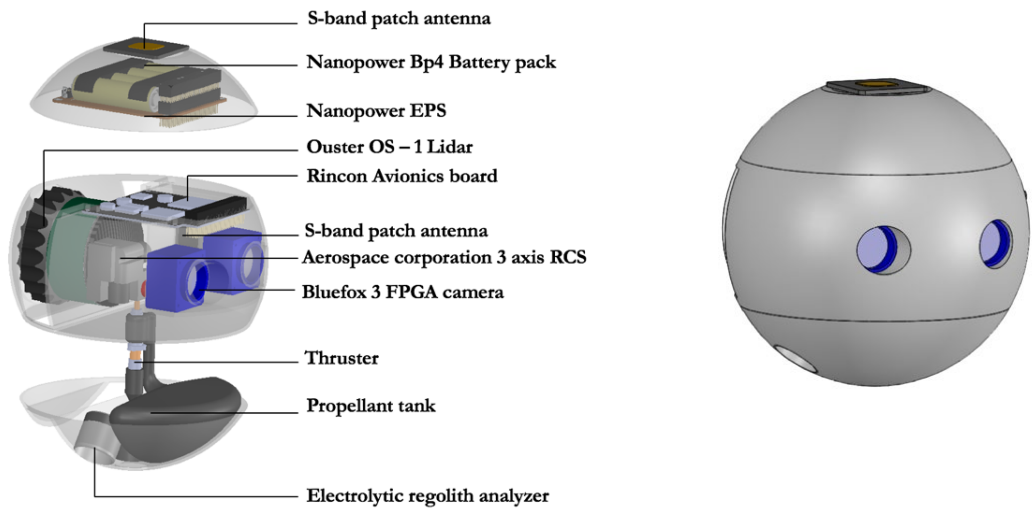

**Figure 6:** Internal (left) and external (right) view of the SphereX robot.

Moreover, each of the SphereX consists of an instrument that uses impedance spectroscopy to determine water content, distribution, and phase in planetary regolith. The instrument consisting of 4 electrodes provides a detailed electrical characterization of the regolith, which offers the potential of significantly increased sensitivity to water and ice[28]. All materials have unique characteristic responses to electromagnetic stimulation that can be used to identify them.

**Remnant Magnetic Field Exploration**

The CubeSat lander to explore the Remnant Magnetic Field at Riener Gamma is shown in (Figure 7) is a 27U custom frame design. This design shown is the result of several design iterations and reviews. There are solar panels surrounding the CubeSat for recharging the batteries on board. The Pumpkin Batteries will power the main onboard computers as well as several slave computers for major subsystems. The lander will use Busek's BGT-5 green monopropellant thrusters. The lander will be an autonomous system guided by the stars using the BCT XCT-50 ADCS. As the lander descends, it will be taking video with its camera and using its Laser Rangefinder for terrain navigation. The lander is expected to make a soft touch-down. The Laser Rangefinder and EMI and dust box is not shown for clarity.

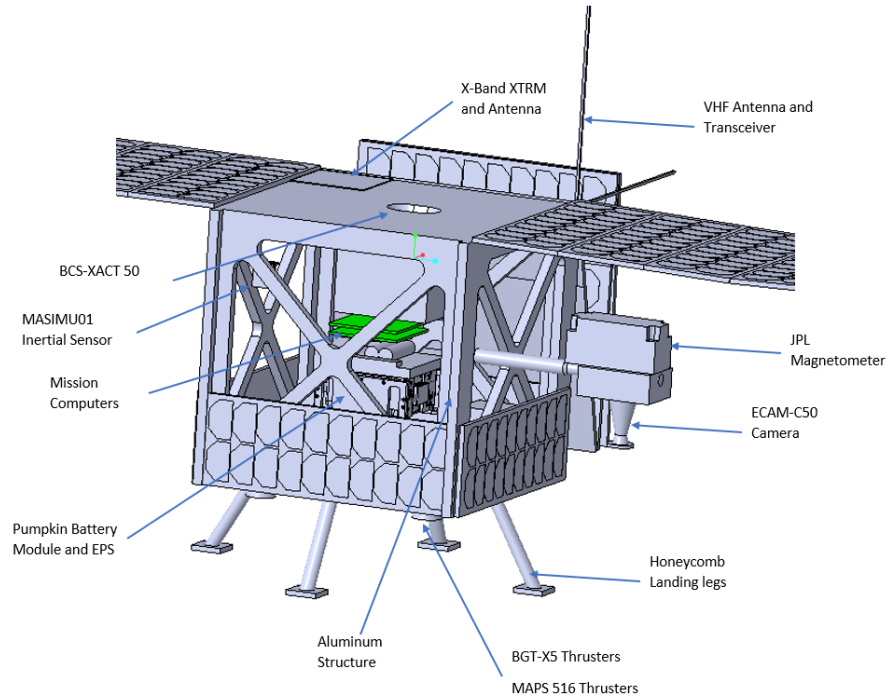

**Figure 7:** Lunar Lander Concept to Explore Remnant Magnetic Fields.

**Table 1:**  Mass and Volume for RMF Lander   **Table 2:** Mass and Volume for Lunar Pits Lander.

| Subsystem | Mass (kg) | Volume (L) | Subsystem | Mass (kg) | Volume (L) |
|---|---|---|---|---|---|
| Propulsion + Landing System | 40.0 | 33.3 | Avionics | 0.13 | 1.1 |
| | | | Power | 1.1 | 0.8 |
| Comms. | 0.5 | 0.4 | ADCS | 0.7 | 0.5 |
| Computer + Electronics | 0.5 | 0.4 | Comms | 0.2 | 0.3 |
| | | | Propulsion | 37.0 | 15 |
| EPS – Battery + Panels | 1.1 | 0.9 | Structure | 2.0 | 0.4 |
| | | | Instruments | 0.4 | 0.6 |
| Instruments | 0.5 | 0.4 | SphereX | 4.4 | 9.1 |
| Structures | 3.2 | 1.2 | Total | 45.9 | 27.8 |
| Margin | 8.1 | 6.4 | Margin | 8.1 | 15 |

The proposed CubeSat lander is within the allowable volume of the 27U CubeSat standard. The lander includes retractable landing legs and a foldout solar panel (Figure 7). In addition, the color camera is deployed to look downwards but is well away from the rocket thrusters. See Table 1 for system budgets. The selected science instruments include JPL's compact magnetometer installed on the Inspire CubeSats and to be used on the Europa Clipper mission. The second is the Malin Ecam-50 which is a 5 Megapixel color camera with heritage from the OSIRIS-REx mission. Both instruments will be used concurrently and be operating as the spacecraft lands in Reiner Gamma and remains there until power is depleted.

**CONCEPT OF OPERATIONS**

**Lunar Pit Exploration**

Figure 8 shows the first phase of the mission. The CubeSat will be launched on NASA's deep space rocket, the Space Launch System (SLS) where it will be encapsulated, docking with the Lunar Gateway presumed to be located at the L2 Lagrange point.

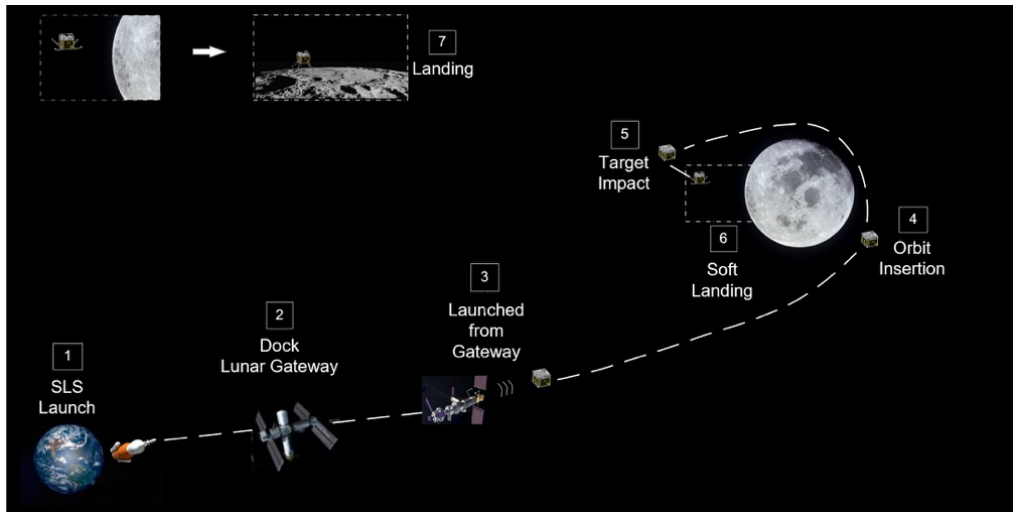

**Figure 8:** Concepts of Operation for the lander.

The lander will be stored on the outside of the Gateway logistics module. The logistics module will be utilized for science including deploying science-focused CubeSat lunar landers. The lander will be in a hibernated state until its ready for science deployment. When ready the CubeSat will be loaded into a P-Pod and launched on its way to the Lunar insertion orbit. The lander has its own propulsion, attitude-determination and control system and communication system. The lander has a total delta-v of 2.5 km/s using its High-Performance Green Propulsion (HPGP) system.

After the lander performs its lunar orbit insertion maneuver, it will perform another impulsive burn to target the landing site. On its way to the landing site, at about 25 km from the surface it will prepare to land. The onboard camera and lens system will be used to navigate and the propulsion system to perform soft landing on the landing site. The CubeSat lander upon successfully landing will deploy the three SphereX robots (Figure 9) using a spring deployment system. The SphereX robots will then hop near the pit entrance, enter the pit and start mapping using the onboard 3D LiDAR sensors and stereo cameras.

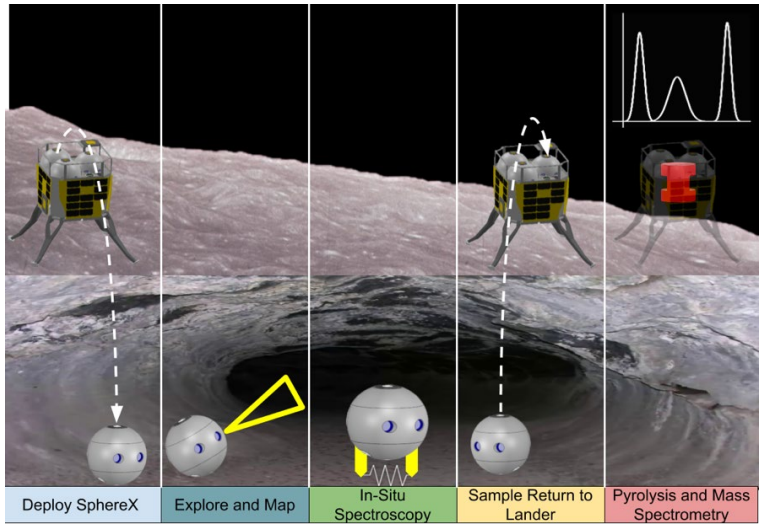

Each robot will also take electrical impedance spectroscopy measurements to determine the water content, distribution, and phase in the planetary regolith inside. Each robot will also collect a few grams of regolith sample and then hop back to the lander. The collected samples will then be inserted inside the VAPoR instrument to perform pyrolysis and mass spectrometry. The CubeSat lander will then transmit the data (electrical impedance spectroscopy, pyrolysis and mass spectroscopy, 3D point cloud data of the pit, images and videos of the landing process, and images and videos of the robots hopping inside the pit) back to the Lunar Gateway. Estimated mission length is 3 days. The lander will land during daytime and there are no plans for the lander to survive the lunar night which simplifies the mission.

**Figure 9:** Concepts of Operation for SphereX.

**Navigation Camera**

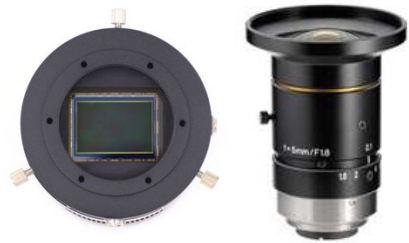

A camera and a lens system will be used for navigating the lander to its landing site. A high-resolution wide-angle camera system (Figure 10) is selected to allow for a single landing navigation camera to minimize subsystem mass and volume. 0.55 meter per pixel imagery of the landing site is available from the Lunar Reconnaissance Orbiter Camera (LROC). Our selected camera system provides 595 meter per pixel resolution at an altitude of 25 km, with a total coverage of over 21,000 square kilometers of lunar surface at that same altitude. At 1 km altitude the camera resolution will be 0.95 meter per pixel. Assuming a uniform distribution of craters, we should expect roughly one 5-20 kilometer crater every 1000 square kilometers, providing an estimated 21 large craters to be used for visual odometry and navigation at 25 km. During landing, the navigation camera will stream imagery to the FPGA and match with the preloaded features, providing position and velocity estimate of the craft. The position estimate is estimated to be three or-ders of magnitude better than current inertial estimates[29]. The expected target deviation from the landing site is expected to be roughly one meter according to studies of Mars Science Laboratory (MSL) touchdown imagery [29].

**Figure 10:** (Left) QHY367C cooled full frame CMOS color camera, (Right) Edmund Optics 5mm FL ultra-high resolution fixed focal length lens.

**Remnant Magnetic Field Exploration**

Figure 11 shows the first phase of the mission, while Figure 12 shows the second phase. The CubeSat will be launched on a Delta IV rocket where it will be encapsulated, docking with the Lunar Gateway presumed to be located at the L2 Lagrange point. The lander will be stored on the outside of the Gateway logistics module. The logistics module will be utilized for science including deploying science-focused CubeSat lunar landers. The lander will be in a hibernated state until its ready for science deployment. When ready the CubeSat will be loaded into a P-Pod and launched on its way to the peri-lune Lunar orbit.

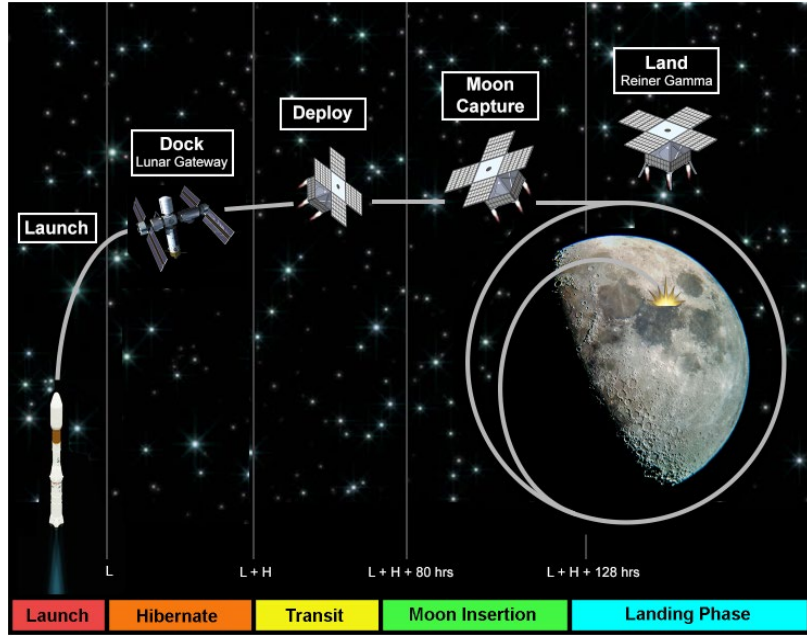

**Figure 11**: Concept of Operations for TCL's Lander.

The lander has its own propulsion, attitude-determination and control system and communication system. The lander may be controlled from the Lunar Gateway or Earth as the onboard radio is DSN compatible. The lander has a total delta-v of 2.5 km/s using its green mono-prop system.

After the lander has been inserted into the perilune orbit, at about 30 km above the surface, the CubeSat will prepare to land. It will rotate 180 degrees where the thrusters will do a strategic burn to slow the CubeSat down, and allow it to be further pulled by lunar gravity, as shown in Figure 12. The controlled descent will take approximately 4-6 minutes culminating in a soft-landing on the surface. While descending, the lander will be taking video at 30 fps and magnetometer readings at 100 Hz. The CubeSat, upon successful landing will radio photos of the touch-down. The CubeSat will continue to take approximately 6 hours of magnetic readings and video footage of the surface. Then it will downlink the data that was recorded back to the Lunar Gateway. Estimated total mission length is 6 days. The lander will avoid landing during the lunar night and there are no plans for the lander to survive the lunar night, which further simplifies the mission.

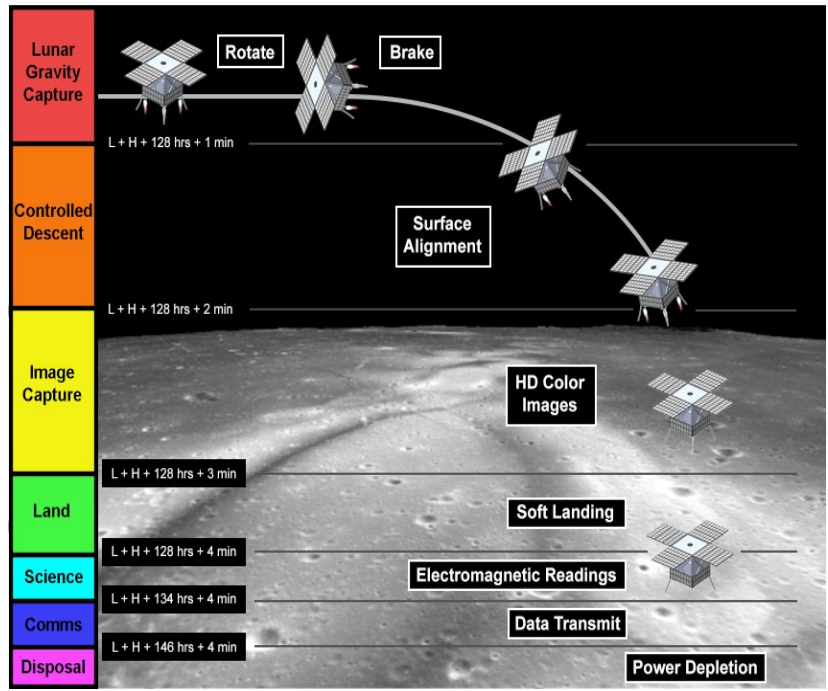

**Figure 12:** TCL Lander's lunar landing sequence.

**DISCUSSIONS**

Presentation of the two mission concepts show the need for advancement in several areas of GNC to enable science exploration. The need for visual navigation to perform pinpoint landing with accuracy of 2 km to 500 meters is an important challenge. The potential opportunities will allow for routine but short missions to the surface of the Moon to obtain samples, perform in-situ analysis and setup instruments. The technology to perform these feats are already there from previous lunar surface missions. The technology does not require the spacecraft be entirely autonomous. With the gateway being nearby, it is possible to partially teleoperate these crafts to minimize on mission operational complexity.

Nevertheless, it is the miniaturization of this technology and use of COTS parts that will be new. This includes advancement in smart-navigation cameras, altitude sensing radars, miniature LIDAR and flash-cameras to name a few. Such technology will be critical to perform the required adjustments during soft-landing on the lunar surface. Apart from the advancement in smart-sensing, we have introduced landers that can deploy from the 27U CubeSat form factor. This is another critical technology advancement. The lander needs to withstand hard landings, not tip over and be upright for positioning communications antennas and navigation cameras.

Technologies we have omitted to keep the mission simple includes the required thermal and power technologies to keep a small lander alive during the lunar night. This maybe a requirement for missions that need to deploy permanent surface instruments. In other cases, the 6-12 days of mission time on the lunar surface appears enough to obtain the required science data.

Overall the proposed Lunar Gateway offers a credible stepping stone to perform exploration of the Moon. CubeSats while still being at their infancy could be used to routinely explore and access all parts of the Moon thus avoiding the risks and high-costs associated with directly sending astronauts. Just as the ISS and its vicinity has become a proving ground in Low Earth Orbit (LEO), the setup also provides an ideal proving ground to advance deep-space technology, terrain navigation technology and technology to perform pin-point landing on the Moon. All of these technologies need to advance for us to realize permanent bases on Mars and the asteroids.

**CONCLUSIONS**

The Lunar Gateway is expected to be positioned on-orbit around the Moon or in a Halo orbit at the L2 Lagrange point. CubeSats are being advanced with onboard propulsion system that provide 500 m/s delta-v utilizing green mono-propellants. This presents limitations for longer interplanetary missions. The trick is to achieve the right trajectory to make possible low-delta-v solutions, but often at the cost of an extended mission. The proposed Lunar Gateway is a game-changer for enabling new science utilizing CubeSats. Deployment of CubeSats from the Lunar Gateway presents a refreshing new opportunity for utilization of these small spacecraft as explorers. In this paper, we outline opportunities for using CubeSats to perform surface science on the Moon and analyze the preliminary feasibility of several potential missions with respect to GNC and propulsion. These new mission opportunities require advancement in terrain navigation and precision landing using COTs technology.

# REFERENCES

[1] R. Zubrin, "Moon Direct: A Cost-Effective Plan to Enable Lunar Exploration and Development," AIAA SciTech 2019 Forum, 2019, San Diego, CA.

[2] J. Schoolcraft, A. Klesh, and T. Werne, "MarCO: Interplanetary Mission Development On a CubeSat Scale," AIAA SpaceOps 2016 Conference, 2016, 2491.

[3] R. Staehle, et al., (2013) "Interplanetary CubeSats: Opening the Solar System to a Broad Community at Lower Cost," Journal of Small Satellites, Vol 2, No. 1, pp. 161-186.

[4] A. Babuscia, T. Choi, KM Cheung, J. Thangavelautham, M. Ravichandran, A. Chandra "Inflatable antenna for CubeSat: Extension of the previously developed S-Band design to the X-Band," AIAA Space 2015 Conference, 4654.

[5] R. Pothamsetti and J. Thangavelautham, "Photovoltaic electrolysis propulsion system for interplanetary CubeSats," 2016 IEEE Aerospace Conference, Big Sky, MT, 2016, pp. 1-10.

[6] National Academies of Sciences, Engineering, and Medicine. 2016. Achieving Science with CubeSats: Thinking Inside the Box. Washington, DC: The National Academies Press.

[7] National Research Council. 2011. Vision and Voyages for Planetary Science in the Decade 2013-2022, National Academies Press.

[8] M.S. Robinson, J. Thangavelautham, R. Wagner, V. Hernandez, J. Finch, "Arne - Exploring the Mare Tranquillitatis Pit," American Geophysical Union Fall Meeting, 2014.

[9] L. Hood, P. Coleman, D. Wilhelms, (1979). "The Moon: Sources of the crustal magnetic anomalies". Science. Vol. 204, pp. 53–57.

[10] D. Blewett, E. Coman, B. Hawke et al. (2011) "Lunar swirls: Examining crustal magnetic anomalies and space weathering trends" Journal of Geophysical Research. Vol. 116.

[11] P. Spudis, "Bubble Bubble - Swirl and Trouble," Smithsonian Air and Space Magazine, 2012.

[12] J. Deca, A. Divin, C. Lue, T. Ahmadi, M. Horanyi, (2018). "Reiner Gamma albedo features reproduced by modeling solar wind standoff," Communications Physics, No. 12

[13] J. Haruyama, et. al., (2009) "Possible lunar lava tube skylight observed by SELENE cameras," Geophysics Research Letters 36, L21206.

[14] M. Robinson, et al., (2012) "Confirmation of sublunarean voids and this layering in mare deposits," Planetary and Space Science 69, pp. 18-27.

[15] R. Wagner, M.Robinson, (2014) "Distribution, formation mechanisms, and significance of lunar pits," Icarus 237 pp. 53-60.

[16] A. Klesh, "INSPIRE and Beyond - Deep Space CubeSats at JPL", 2015.

[17] C. Taylor, A. Shao, N. Armade et al., (2013) "Hummingbird: Versatile Interplanetary Mission Architecture," Interplanetary Small Satellite Conference.

[18] C. Hardgrove, J. Bell, J. Thangavelautham et al., "The Lunar Polar Hydrogen Mapper (LunaH-Map) mission: Mapping hydrogen distributions in permanently shadowed regions of the Moon's south pole", 46th LPSC, 2015.

[19] P. Clark, B. Malphrus, K. Brown, "Lunar Ice Cube Mission: Determining Lunar Water Dynamics with a First Generation Deep Space CubeSat," 47th Lunar and Planetary Science Conference, 2016.

[20] U. Wehmeier et al., "The Lunar Flashlight CubeSat instrument: a compact SWIR laser reflectometer to quantify and map water ice on the surface of the moon,"Proc. SPIE 10769, CubeSats and NanoSats for Remote Sensing II, 107690H.

[21] M. Robinson, J. Thangavelautham, B. Anderson, et al. "Swirl: Unravelling an Enigma," Planetary and Space Sciences Special Issue, pp. 1-33, 2018.

[22] L. McNutt, L. Johnson, P. Kahn, et al., "Near-Earth Asteroid (NEA) Scout," AIAA SPACE, 2014.

[23] T. Hashimoto, "The World's Smallest Moon Lander, OMOTENASHI", JAXA, 2017.

[24] F. Topputo, M. Massari, J. Bigg et al. "LUMIO: Lunar Meteoroid Impact Observer" ICubeSat Conference 2017.

[25] J. Thangavelautham, M. S. Robinson, A. Taits, et al., "Flying, hopping Pit-Bots for cave and lava tube exploration on the Moon and Mars" *2nd International Workshop on Instrumentation for Planetary Missions*, NASA Goddard, 2014.
12